\documentclass[a4paper,11pt]{article}
\pdfoutput=1 

\usepackage{jheppub} 

\usepackage[utf8]{inputenc} 
\usepackage[T1]{fontenc} 
\usepackage{xcolor}
\allowdisplaybreaks

\makeatletter
\def\@fpheader{\relax}
\makeatother
\title{\boldmath Renormalization of a model for spin-1 matter fields}

\author[a,b]{Ailier Rivero-Acosta}
\author[a,c]{and Carlos A. Vaquera-Araujo}

\affiliation[a]{Departamento de F\'isica, DCI, Campus Le\'on, Universidad de Guanajuato, Loma del Bosque 103, Lomas del Campestre, Le\'on C.P. 37150, Guanajuato, Mexico}
\affiliation[b]{Laboratorio de Ciencia Planetaria, Departamento de F\'isica, Universidad Central de Las Villas, Santa Clara C.P. 54830, Villa Clara, Cuba}
\affiliation[c]{Consejo Nacional de Ciencia y Tecnolog\'ia, Av. Insurgentes Sur 1582, Colonia Cr\'edito Constructor, Del. Benito Ju\'arez, Ciudad de Mexico C.P. 03940, Mexico}

\emailAdd{ailierrivero@gmail.com}
\emailAdd{vaquera@fisica.ugto.mx}

\abstract{In this work, the one-loop renormalization of a theory for fields transforming in the $(1,0)\oplus(0,1)$ representation of the Homogeneous Lorentz Group is studied.  The model includes an arbitrary gyromagnetic factor and  self-interactions of the spin 1 field, which has mass dimension one. The model is shown to be renormalizable for any value of the gyromagnetic factor.}

\begin{document} 
\maketitle
\flushbottom
\section{Introduction}

In the Standard Model of particle physics, only fields transforming in the $(0,0)$, $(1/2,0)$, $(0,1/2)$ and $(1/2,1/2)$ representations of the Homogeneous Lorentz Group (HLG) are needed. There is however no guiding principle restricting the possible representations, and indeed high spin fields naturally appear in Hadron physics and in Beyond the Standard Model (BSM) scenarios like supergravity and superstrings.

In an attempt to better understand the physics of fields transforming in different representations of the HLG, a series of  works  have been carried \cite{Napsuciale:2006wr,DelgadoAcosta:2009ic,Napsuciale:2007ry,Delgado-Acosta:2013nia,AngelesMartinez:2011nt,VaqueraAraujo:2012qa,Vaquera-Araujo:2013bwa,Gomez-Avila:2013qaa} based on the projection onto subspaces of the Poincaré group. In this formalism, it has been shown that the gyromagnetic factor of spin 3/2 fields is connected with their causal propagation in an electromagnetic background \cite{Napsuciale:2006wr}, and with the unitarity of the Compton scattering amplitude in the forward direction \cite{DelgadoAcosta:2009ic}. The formalism can also be applied to lower spins, for example, in the spin 1 case, a similar connection between the unitarity of Compton scattering in the forward direction and the gyromagnetic factor of the field exists, which is also related to the electric quadrupole moment \cite{Napsuciale:2007ry}. 

When the Poincaré projector method is applied to spin 1/2 fields transforming  in the $(1/2,0)\oplus(0,1/2)$ representation \cite{VaqueraAraujo:2012qa,Vaquera-Araujo:2013bwa}, the resulting Lagrangian is a generalized version of the original second order Feynman-Gell-Man formalism \cite{Feynman:1958ty}, enhanced with an arbitrary gyromagnetic factor and fermion self interactions. The second order fermions studied in these works are conceptually different to Dirac ones, as the former propagate 8 dynamical degrees of freedom instead of 4. As shown in \cite{VaqueraAraujo:2012qa,Vaquera-Araujo:2013bwa}, there is a consistent reduction of dynamical degrees of freedom and a direct connection between the renormalization group equations for the second order fermions and the Dirac formalism if the gyromagnetic (or chromomagnetic)  factor is set to the fixed value $g = 2$.

The goal of the present work is to study the renormalization properties of  spin-1 matter fields\footnote{Here we understand matter fields as massive non-gauge fields.}  transforming in the $(1,0)\oplus(0,1)$ representation of the HLG in a model based on the Poincaré projector formalism, as a direct generalization of the spin 1/2 case \cite{VaqueraAraujo:2012qa,Vaquera-Araujo:2013bwa}. 

The difference between the pure spin 1 representation $(1,0)\oplus(0,1)$, described by an antisymmetric tensor field of second rank, and the more familiar $(1/2,1/2)$ vector field is more dramatic in the massless case, as the Kalb-Ramond  antisymmetric gauge field contains only one physical longitudinal degree of freedom \cite{Kalb:1974yc}, whereas the massless vector gauge field is characterized by 2 transverse ones. Switching to massive spin-1 particles, one must distinguish between gauge invariant and non-gauge invariant theories. It can be shown that a massive Stueckelberg compensated Kalb-Ramond gauge field is dual to a compensated massive gauge vector field \cite{Smailagic:2001ch}. However, for non-gauge invariant massive spin-1 theories, the properties of four-vector and antisymmetric tensor particles can differ significantly.  In \cite{Chizhov:2011zz} the difference between spin-1 antisymmetric tensor mesons and the four-vector mesons has been studied
in detail for composite hadrons. In the present work, we focus instead on pointlike massive spin-1 bosons, with emphasis on their electromagnetic properties and their possible self-interactions. 

The model studied here is based on \cite{Delgado-Acosta:2013nia}, where the complex antisymmetric tensor field has 6 complex degrees of freedom, making the $(1,0)\oplus(0,1)$ theory explicitly different to any of a massive gauge vector field. In  \cite{Delgado-Acosta:2013nia} the Compton scattering of spin-1 particles described by both a massive four-vector and an antisymmetric tensor was analyzed for arbitrary values of the gyromagnetic factor, finding that the Compton scattering cross section off the parity degrees of freedom in $(1,0)\oplus(0,1)$ is finite in the forward direction, though it is still divergent
elsewhere. Interestingly, for the antisymmetric tensor this result is independent of the gyromagnetic factor, while Compton scattering off the four-vector is only well
behaved in all directions provided the gyromagnetic ratio is set to $g=2$.  Given the non-finiteness of Compton scattering in this theory, it is unclear if the renormalizable theory described here corresponds to a perturbation theory about a sensible zeroth-order Hamiltonian. However, it constitutes a unique theoretical laboratory from the point of view of the renormalization group, in the same spirit as scalar $\lambda \phi^3$ theory.

The structure of the paper is the following: In section \ref{model} we describe the model and the Feynman rules. The renormalization procedure is presented in section \ref{reno} together with the cancellation of all the potentially divergent contributions to the one-loop vertices of the theory. Finally, the conclusions of the work are discussed in section \ref{conc}.

\section{The Model}\label{model}
Our model comprises a massive complex spin-1 antisymmetric tensor field $B^{\alpha\beta}$ in the $(1,0)\oplus(0,1)$ representation of the HLG, minimally coupled to $U(1)_{\mathrm{EM}}$ with arbitrary gyromagnetic factor and mass dimension one, allowing for self interaction terms. The Lagrangian of the model is given by
\begin{eqnarray}
&&\mathcal {L}=-{\frac {1}{4}}F^{\mu \nu }F_{\mu \nu }+ (D^{\mu}B^{\alpha\beta})^{\dagger} \left(T_{\mu\nu}\right)_{\alpha\beta\gamma\delta} (D^{\nu}B^{\gamma\delta}) -m^2 {(B^{\alpha\beta})}^{\dagger}B_{\alpha\beta}\nonumber\\
&&+\frac{\lambda_{1}}{2}(B^{\alpha\beta\,\dagger}1_{\alpha\beta\gamma\delta}B^{\gamma\delta})(B^{\mu\nu\,\dagger}1_{\mu\nu\rho\sigma}B^{\rho\sigma})+\frac{\lambda_{2}}{2}(B^{\alpha\beta\,\dagger}\chi_{\alpha\beta\gamma\delta}B^{\gamma\delta})(B^{\mu\nu\,\dagger}\chi_{\mu\nu\rho\sigma}B^{\rho\sigma})\nonumber\\
&&+\frac{\lambda_{3}}{2}(B^{\alpha_{1}\beta_{1}\,\dagger}(M^{\mu\nu})_{\alpha_{1}\beta_{1}\gamma_{1}\delta_{1}}B^{\gamma_{1}\delta_{1}})(B^{\alpha_{2}\beta_{2}\,\dagger}(M_{\mu\nu})_{\alpha_{2}\beta_{2}\gamma_{2}\delta_{2}}B^{\gamma_{2}\delta_{2}})\nonumber\\
&&+\frac{\lambda_{4}}{2}(B^{\alpha_{1}\beta_{1}\,\dagger}(S^{\mu\nu})_{\alpha_{1}\beta_{1}\gamma_{1}\delta_{1}}B^{\gamma_{1}\delta_{1}})(B^{\alpha_{2}\beta_{2}\,\dagger}(S_{\mu\nu})_{\alpha_{2}\beta_{2}\gamma_{2}\delta_{2}}B^{\gamma_{2}\delta_{2}}),
\end{eqnarray}
where $D^{\mu}=\partial^{\mu}+ieA^{\mu}$ is the covariant derivative, and the tensors used are given by
\begin{eqnarray}
&&F^{\mu \nu}=\partial^{\mu}A^{\nu}-\partial^{\nu}A^{\mu},\quad\, T_{\mu\nu}=g_{\mu\nu}1_{\alpha\beta\gamma\delta}-ig(M_{\mu\nu})_{\alpha\beta\gamma\delta},\nonumber\\
&&1_{\alpha\beta\gamma\delta}=\frac{1}{2}(g_{\alpha\gamma}g_{\beta\delta}-g_{\alpha\delta}g_{\beta\gamma}),\quad\, \chi_{\alpha\beta\gamma\delta}=\frac{i}{2}\epsilon_{\alpha\beta\gamma\delta},\nonumber\\
&&(M_{\mu\nu})_{\alpha\beta\gamma\delta}=-i(g_{\mu\gamma}1_{\alpha\beta\nu\delta}+g_{\mu\delta}1_{\alpha\beta\gamma\nu}-g_{\gamma\nu}1_{\alpha\beta\mu\delta}-g_{\delta\nu}1_{\alpha\beta\gamma\mu}),\nonumber\\
&& (S_{\mu\nu})_{\alpha\beta\gamma\delta}=g_{\mu\nu}1_{\alpha\beta\gamma\delta}-g_{\mu\gamma}1_{\alpha\beta\nu\delta}-g_{\mu\delta}1_{\alpha\beta\gamma\nu}-g_{\gamma\nu}1_{\alpha\beta\mu\delta}-g_{\delta\nu}1_{\alpha\beta\gamma\mu}.
\end{eqnarray}
The kinetic part of the Lagrangian is of Klein-Gordon type and spin-1 information is encoded by a Pauli-like term modulated by an arbitrary gyromagnetic factor $g$ and the four independent quartic self-interactions that can be built from the covariant basis for the $(1,0)\oplus(0,1)$ representation space, given by the complete set of tensors presented in \cite{Gomez-Avila:2013qaa}, namely $\{1,\chi, M^{\mu\nu},  S^{\mu\nu}, \chi S^{\mu\nu},   C^{\mu\nu\alpha\beta}\}$ with
\begin{equation}
C_{\mu\nu\alpha\beta}=4\{M_{\mu\nu},M_{\alpha\beta}\}+2\{M_{\mu\alpha},M_{\nu\beta}\}-2\{M_{\mu\beta},M_{\nu\alpha}\}-16 (1_{\mu\nu\alpha\beta}).
\end{equation}

\begin{figure}[h]
\centering 
\includegraphics[width=\textwidth]{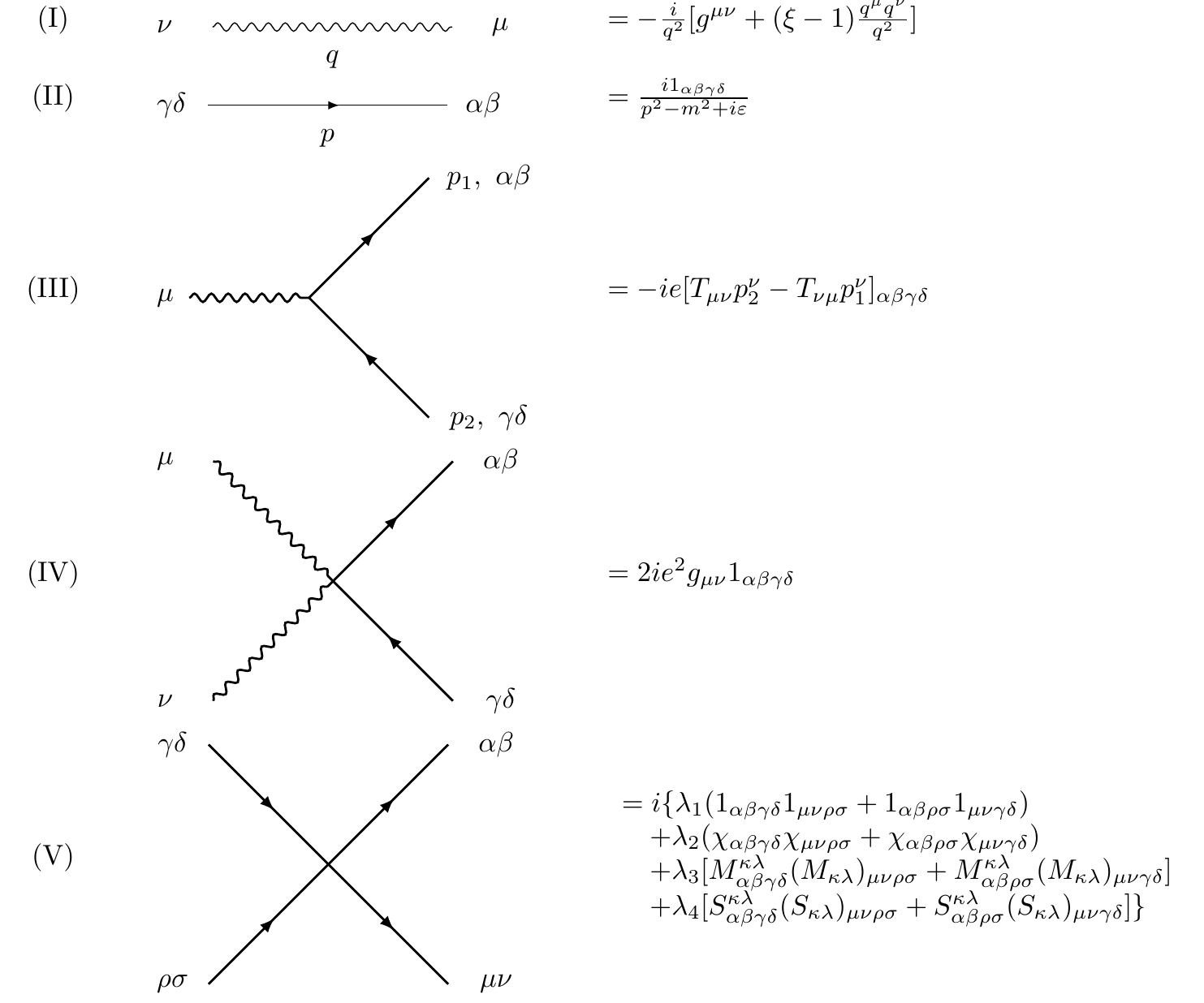}
\caption{\label{fig:FR} Feynman rules of the model.}
\end{figure}

In our analysis, the gauge freedom is fixed by the $R_{\xi}$ contribution
\begin{equation}
\mathcal {L}_{\mathrm{G.F.}}=-\frac{1}{2\xi}(\partial^{\mu}A_{\mu})^2
\end{equation}
with arbitrary gauge fixing parameter $\xi$, rendering the complete Lagrangian of the model as\begin{eqnarray}
\mathcal {L}&=&-{\frac {1}{4}}F^{\mu \nu }F_{\mu \nu }-\frac{1}{2\xi}(\partial^{\mu}A_{\mu})^2+\partial^{\mu}B^{\alpha\beta\dagger}\partial_{\mu}B_{\alpha\beta}-m^2 {(B^{\alpha\beta})^{\dagger}}B_{\alpha\beta}\nonumber\\
&&-ieA^{\mu}[B^{\alpha\beta\dagger}(T_{\mu\nu})_{\alpha\beta\gamma\delta}\partial^{\nu}B^{\gamma\delta}-(\partial^{\nu}B^{\alpha\beta\dagger})(T_{\nu\mu})_{\alpha\beta\gamma\delta}B^{\gamma\delta}]+e^2A^{\mu}A_{\mu}B^{\alpha\beta\dagger}B_{\alpha\beta} \nonumber\\
&&+\frac{\lambda_{1}}{2}(B^{\alpha\beta\,\dagger}1_{\alpha\beta\gamma\delta}B^{\gamma\delta})(B^{\mu\nu\,\dagger}1_{\mu\nu\rho\sigma}B^{\rho\sigma})+\frac{\lambda_{2}}{2}(B^{\alpha\beta\,\dagger}\chi_{\alpha\beta\gamma\delta}B^{\gamma\delta})(B^{\mu\nu\,\dagger}\chi_{\mu\nu\rho\sigma}B^{\rho\sigma})\nonumber\\
&&+\frac{\lambda_{3}}{2}(B^{\alpha_{1}\beta_{1}\,\dagger}(M^{\mu\nu})_{\alpha_{1}\beta_{1}\gamma_{1}\delta_{1}}B^{\gamma_{1}\delta_{1}})(B^{\alpha_{2}\beta_{2}\,\dagger}(M_{\mu\nu})_{\alpha_{2}\beta_{2}\gamma_{2}\delta_{2}}B^{\gamma_{2}\delta_{2}})\nonumber\\
&&+\frac{\lambda_{4}}{2}(B^{\alpha_{1}\beta_{1}\,\dagger}(S^{\mu\nu})_{\alpha_{1}\beta_{1}\gamma_{1}\delta_{1}}B^{\gamma_{1}\delta_{1}})(B^{\alpha_{2}\beta_{2}\,\dagger}(S_{\mu\nu})_{\alpha_{2}\beta_{2}\gamma_{2}\delta_{2}}B^{\gamma_{2}\delta_{2}}).
\label{LagB}
\end{eqnarray}
The Feynman rules corresponding to the  above Lagrangian  are presented in Fig.  \ref{fig:FR}, where all momenta are incoming.

The gauge invariance of the theory imposes two important Ward-Takahashi identities (see \cite{AngelesMartinez:2011nt} for their derivation in the analogous spin 1/2 case). The first one relates the tensor-tensor-photon (TT$\gamma$) vertex function $-ie\Gamma^\mu(q,p,-p-q)$, where $q$ is the momentum of the photon, with the tensor self-energy $-i\Sigma(p)$ according to 
\begin{equation}\label{Ward1}
\Gamma^\mu(0,p,-p)=-\frac{\partial\Sigma(p)}{\partial p_\mu}.
\end{equation}
The second one involves the tensor-tensor-photon-photon (TT$\gamma$$\gamma$) vertex $ie^2\Gamma^{\mu\nu}(q,q',p,p')$, with photon momenta $q$ and $q'$, and the TT$\gamma$ vertex, and reads
\begin{equation}\label{Ward2}
\Gamma^{\mu\nu}(0,q',p,p')=\frac{\partial\Gamma_\nu(q',p,p')}{\partial p_\mu}+\frac{\partial\Gamma_\nu(q',p,p')}{\partial p'_\mu}.
\end{equation}

\section{Renormalization}\label{reno}
In this section, we analyze the renormalization properties of the model at one-loop level, studying the UV divergent parts of all the potentially divergent vertex functions. In this work, we use dimensional regularization with $d=4-2\epsilon$ and the naive prescription for the chirality operator $\chi$
\begin{equation}
[\chi,M^{\mu\nu}]=0,\qquad \{\chi,S^{\mu\nu}\}=0.
\end{equation}
This approach does not lead to inconsistencies as $\chi$ appears in pairs for all the processes involved. 
The subtraction scheme used in the study is the minimal subtraction (MS) one.
\subsection{Counterterms}
\begin{figure}[h]
\centering 
\includegraphics[width=\textwidth]{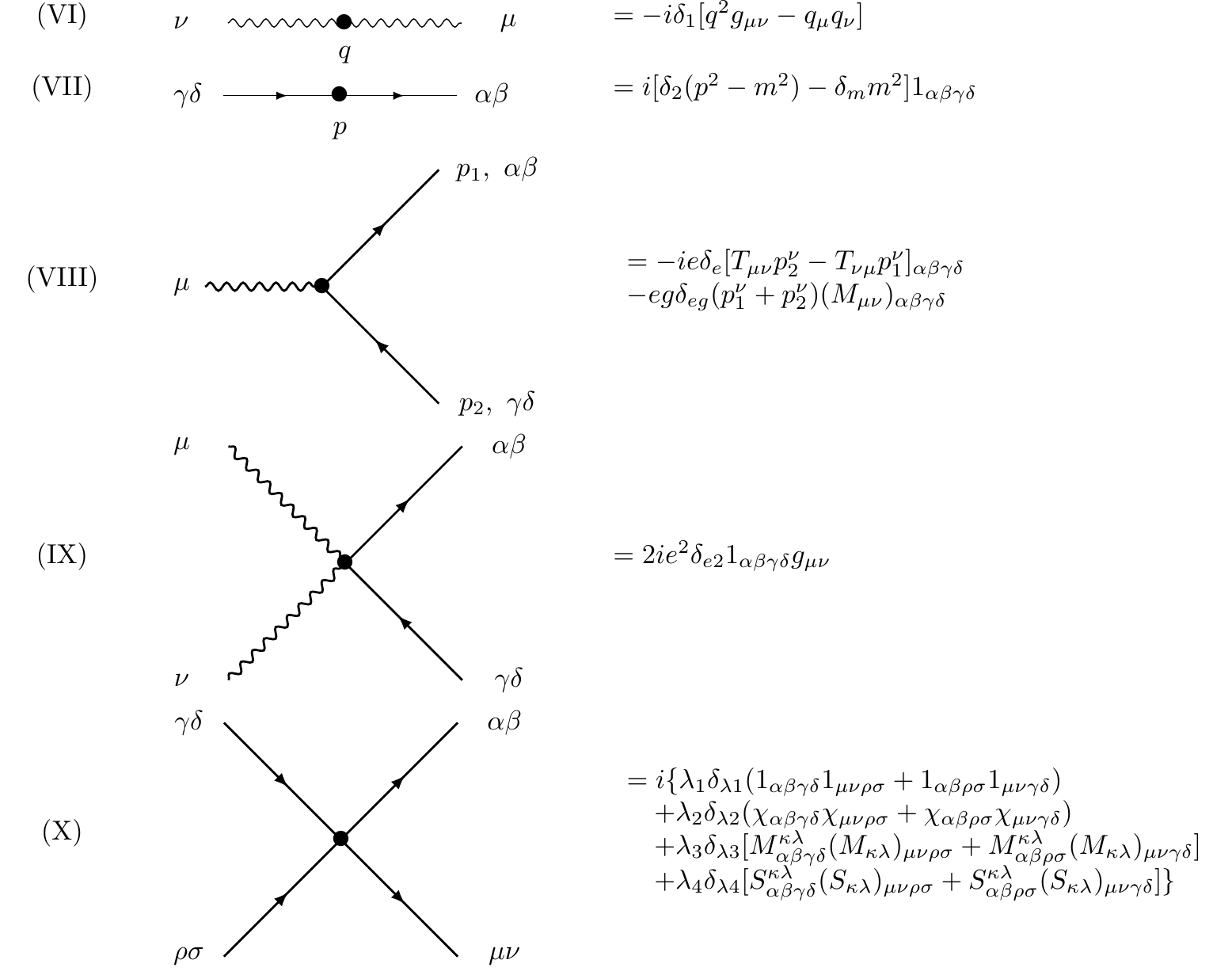}
\caption{\label{fig:FR2} Feynman rules for the counterterms.}
\end{figure}

Taking Eq.(\ref{LagB}) as the bare Lagrangian, with all bare quantities denoted by a $0$ subscript, its parameters are the tensor mass $m_0$, the tensor charge $e_0$ and the
gyromagnetic factor $g_0$. The renormalized fields are defined in terms of the bare ones through
\begin{equation}
A^{\mu}_{r}=Z_{1}^{-\frac{1}{2}}A^{\mu}_{0},\,\,\,\,\,\,\,\,\,\,\,B^{\alpha\beta}_{r}=Z_{2}^{-\frac{1}{2}}B^{\alpha\beta}_{0}.
\end{equation}
It is convenient to split the Lagrangian as the sum of two terms 
\begin{equation}
\mathcal {L}_0=\mathcal {L}_r+\mathcal {L}_{ct},
\end{equation}
where the first piece is the renormalized Lagrangian, and has the same structure as Eq.(\ref{LagB})
\begin{eqnarray}
\mathcal {L}_{r}&=&-{\frac {1}{4}}F^{\mu \nu }_{r}F_{r\,\mu \nu }-\frac{1}{2\xi_r}(\partial_{\mu}A_{r}^{\mu})^2+\partial^{\mu}B^{\alpha\beta\dagger}_{r}\partial_{\mu}B_{r\,\alpha\beta}-m_{r}^2 {(B^{\alpha\beta}_{r})^{\dagger}}B_{r\,\alpha\beta}\\\nonumber
&&-ie_{r} A^{\mu}_{r}[B^{\alpha\beta\dagger}_{r}(T_{r\,\mu\nu})_{\alpha\beta\gamma\delta}\partial^{\nu}B^{\gamma\delta}_{r}-(\partial^{\nu}B^{\alpha\beta\dagger}_{r})(T_{r\,\nu\mu})_{\alpha\beta\gamma\delta})B^{\gamma\delta}_{r}]\\\nonumber
&&+e_{r}^2 A^{\mu}_{r}A_{r\,\mu}B^{\alpha\beta\dagger}_{r}B_{r\,\alpha\beta}+\frac{\lambda_{r1}}{2}(B^{\alpha\beta\,\dagger}_{r}1_{\alpha\beta\gamma\delta}B^{\gamma\delta}_{r})(B^{\mu\nu\,\dagger}_{r}1_{\mu\nu\rho\sigma}B^{\rho\sigma}_{r})\\\nonumber
&&+\frac{\lambda_{r2}}{2}(B^{\alpha\beta\,\dagger}_{r}\chi_{\alpha\beta\gamma\delta}B^{\gamma\delta}_{r})(B^{\mu\nu\,\dagger}_{r}\chi_{\mu\nu\rho\sigma}B^{\rho\sigma}_{r})\\\nonumber
&&+\frac{\lambda_{r3}}{2}(B^{\alpha_{1}\beta_{1}\,\dagger}_{r}(M^{\mu\nu})_{\alpha_{1}\beta_{1}\gamma_{1}\delta_{1}}B^{\gamma_{1}\delta_{1}}_{r})(B^{\alpha_{2}\beta_{2}\,\dagger}_{r}(M_{\mu\nu})_{\alpha_{2}\beta_{2}\gamma_{2}\delta_{2}}B^{\gamma_{2}\delta_{2}}_{r})\\\nonumber
&&+\frac{\lambda_{r4}}{2}(B^{\alpha_{1}\beta_{1}\,\dagger}_{r}(S^{\mu\nu})_{\alpha_{1}\beta_{1}\gamma_{1}\delta_{1}}B^{\gamma_{1}\delta_{1}}_{r})(B^{\alpha_{2}\beta_{2}\,\dagger}_{r}(S_{\mu\nu})_{\alpha_{2}\beta_{2}\gamma_{2}\delta_{2}}B^{\gamma_{2}\delta_{2}}_{r}),
\label{EqL1}
\end{eqnarray}
and the second one contains the relevant counterterms
\begin{eqnarray}
\mathcal {L}_{ct}&=&-{\frac {1}{4}}\delta_1 F^{\mu \nu }_{r}F_{r\,\mu \nu }+\delta_2[\partial^{\mu}B^{\alpha\beta\dagger}_{r}\partial_{\mu}B_{r\,\alpha\beta}-m_{r}^2 {(B^{\alpha\beta}_{r})^{\dagger}}B_{r\,\alpha\beta}]-\delta_{m}m_{r}^2 {(B^{\alpha\beta}_{r})^{\dagger}}B_{r\,\alpha\beta}\nonumber\\
&&-ie_{r}\delta_{e} A^{\mu}_{r}[B^{\alpha\beta\dagger}_{r}(T_{r\,\mu\nu})_{\alpha\beta\gamma\delta}\partial^{\nu}B^{\gamma\delta}_{r}-(\partial^{\nu}B^{\alpha\beta\dagger}_{r})(T_{r\,\nu\mu})_{\alpha\beta\gamma\delta}B^{\gamma\delta}_{r}]\nonumber\\
&&-ie_{r}\delta_{eg} A^{\mu}_{r}[B^{\alpha\beta\dagger}_{r}(-ig_{r})(M_{\mu\nu})_{\alpha\beta\gamma\delta}\partial^{\nu}B^{\gamma\delta}_{r}-(\partial^{\nu}B^{\alpha\beta\dagger}_{r})(ig_{r})(M_{\mu\nu})_{\alpha\beta\gamma\delta}B^{\gamma\delta}_{r}]\nonumber\\
&&+\delta_{e2} e_{r}^2 A^{\mu}_{r}A_{r\,\mu}B^{\alpha\beta\dagger}_{r}B_{r\,\alpha\beta}+\frac{\lambda_{r1}}{2}\delta_{\lambda 1}(B^{\alpha\beta\,\dagger}_{r}1_{\alpha\beta\gamma\delta}B^{\gamma\delta}_{r})(B^{\mu\nu\,\dagger}_{r}1_{\mu\nu\rho\sigma}B^{\rho\sigma}_{r})\nonumber\\
&&+\frac{\lambda_{r2}}{2}\delta_{\lambda 2}(B^{\alpha\beta\,\dagger}_{r}\chi_{\alpha\beta\gamma\delta}B^{\gamma\delta}_{r})(B^{\mu\nu\,\dagger}_{r}\chi_{\mu\nu\rho\sigma}B^{\rho\sigma}_{r})\nonumber\\
&&+\frac{\lambda_{r3}}{2}\delta_{\lambda 3}(B^{\alpha_{1}\beta_{1}\,\dagger}_{r}(M^{\mu\nu})_{\alpha_{1}\beta_{1}\gamma_{1}\delta_{1}}B^{\gamma_{1}\delta_{1}}_{r})(B^{\alpha_{2}\beta_{2}\,\dagger}_{r}(M_{\mu\nu})_{\alpha_{2}\beta_{2}\gamma_{2}\delta_{2}}B^{\gamma_{2}\delta_{2}}_{r})\nonumber\\
&&+\frac{\lambda_{r4}}{2}\delta_{\lambda 4}(B^{\alpha_{1}\beta_{1}\,\dagger}_{r}(S^{\mu\nu})_{\alpha_{1}\beta_{1}\gamma_{1}\delta_{1}}B^{\gamma_{1}\delta_{1}}_{r})(B^{\alpha_{2}\beta_{2}\,\dagger}_{r}(S_{\mu\nu})_{\alpha_{2}\beta_{2}\gamma_{2}\delta_{2}}B^{\gamma_{2}\delta_{2}}_{r}),
\label{EqL2}
\end{eqnarray}
with the following definitions 
\begin{equation}
\begin{matrix}
\delta_1\equiv Z_{1}-1,&\quad&\delta_2\equiv Z_{2}-1,&\quad&\delta_{m}\equiv Z_{m}-Z_{2},&\quad& \delta_{e}\equiv Z_{e}-1,\\
\delta_{eg}\equiv Z_{eg}-Z_{e}, &\quad& \delta_{e2}\equiv Z_{e2}-1,&\quad&\delta_{\lambda j}\equiv Z_{\lambda j}-1,&\quad&\xi_r\equiv Z_1^{-1}\xi_0,
\end{matrix}\label{deltas}
\end{equation}
and
\begin{equation}
Z_{m}\equiv \frac{m_{0}^2}{m_{r}^2}Z_{2}, \quad\, Z_{e}\equiv \frac{e_{0}}{e_{r}}Z_{1}^{\frac{1}{2}}Z_{2}, \quad\, Z_{eg}\equiv \frac{g_{0}}{g_{r}}Z_{e}, \quad\, Z_{e2}\equiv \frac{e_{0}^2}{e_{r}^2}Z_{1}Z_{2} \quad\, Z_{\lambda j}\equiv \frac{\lambda_{0 j}}{\lambda_{rj}}{Z_2}^{2}.\label{Zs}
\end{equation}
In $d=4-2\epsilon$ dimensions, the renormalized parameters must be scaled according to
\begin{equation}
e_r\rightarrow \mu^{\epsilon}e_r,\quad g_r\rightarrow g_r,\quad \lambda_{r\, i}\rightarrow \mu^{2\epsilon}\lambda_{r\, i},\quad m_r\rightarrow m_r,
\end{equation}
where $\mu$ is the arbitrary scale introduced by dimensional regularization. In what follows, we will omit the $r$ subscript for the renormalized parameters. In this notation, the Feynman rules for counterterms are given in Fig. \ref{fig:FR2}.
In the following subsections, we will compile the results obtained for the calculation of all the divergent processes showing that all the divergencies can be absorbed successfully into the given set of counterterms provided by the theory.

\subsection{Vacuum Polarization}
\begin{figure}[h]
\centering 
\includegraphics[width=.8\textwidth]{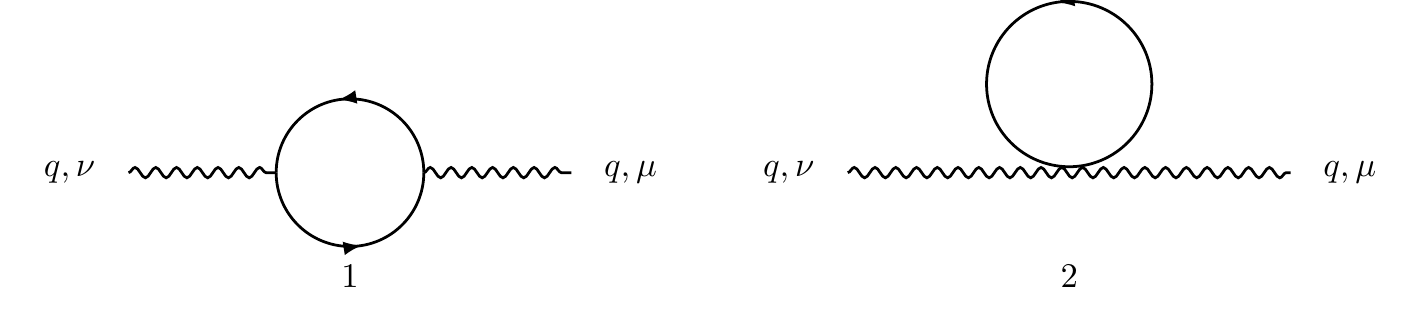}
\caption{\label{fig:ph} Feynman diagrams for the vacuum polarization at one-loop.}
\end{figure}
There are two diagrams contributing to the vacuum polarization, depicted in Figure \ref{fig:ph}. The divergent piece, denoted by $-i\Pi^{\mu\nu}(q)^{*}$ is given by
\begin{equation}
-i\Pi^{\mu\nu}(q)^{*}=i\frac{e^{2}(2g^{2}-1)}{{8\pi^{2}\epsilon}}(q^{2}g^{\mu\nu}-q^{\mu}q^{\nu}),
\end{equation}
and can be removed in the MS scheme by fixing the counterterm $\delta_{1}$ as
\begin{equation}
\delta_{1}=\frac{e^2(2g^{2}-1)}{8\pi^{2}\epsilon}\label{delta1}.
\end{equation}

\subsection{Tensor self-energy}
\begin{figure}[h]
\centering 
\includegraphics[width=.9\textwidth]{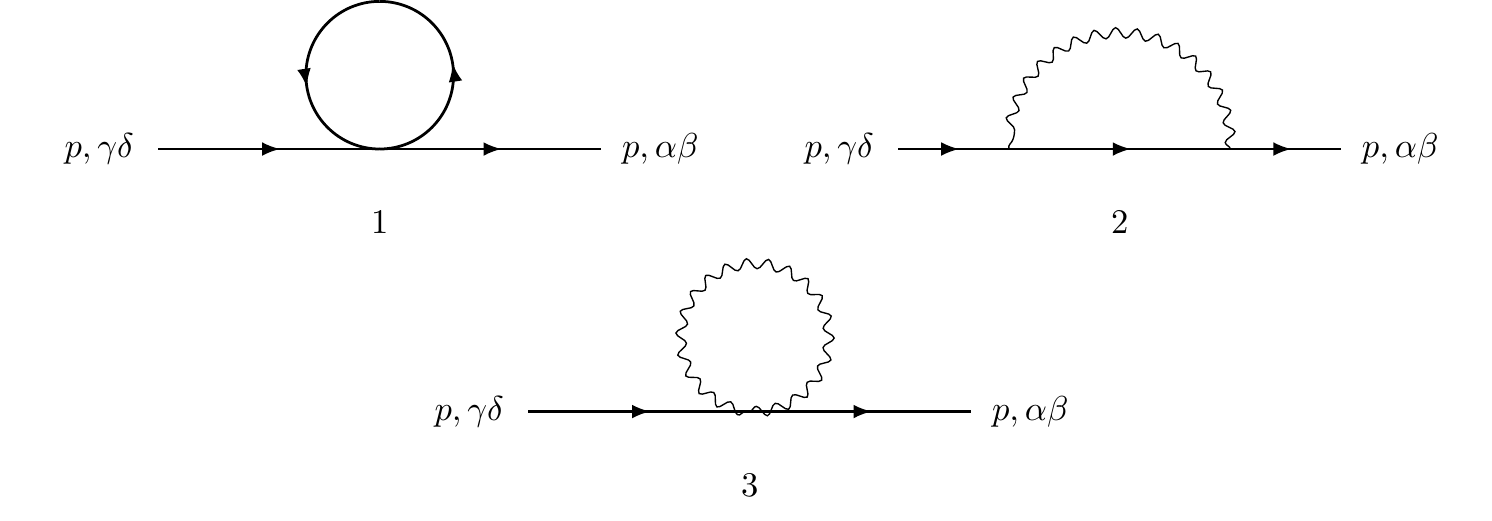}
\caption{\label{fig:ten} Feynman diagrams for the Tensor self-energy at one-loop.}
\end{figure}

In Figure \ref{fig:ten} are shown the three diagrams contributing to the Tensor self-energy. The divergent part of this amplitude is 
\begin{eqnarray}
-i\Sigma^*_{\alpha\beta\gamma\delta}(p)&=&\frac{-i}{32\pi^2 \epsilon}\Big\{m^2 \left(2 e^2 g^2+e^2 \xi +7 \lambda _1+\lambda _2+8 \lambda _3+12\lambda _4\right)\nonumber\\
&&-e^2 (\xi -3){p}^2\Big\}1_{\alpha\beta\gamma\delta},
\end{eqnarray}
and the counterterms that cancel the UV divergence are then given by
\begin{eqnarray}
&&\delta_{2}=-\frac{e^2 (\xi -3)}{16 \pi ^2 \epsilon },\label{delta2}\\
&&\delta_{m}=-\frac{e^2 \left(2 g^2+3\right)+7 \lambda _1+\lambda _2+8 \lambda _3+12\lambda _4}{16 \pi ^2 \epsilon }\label{deltam}.
\end{eqnarray}

\subsection{$\gamma\gamma\gamma$ vertex}
\begin{figure}[h]
\centering 
\includegraphics[width=.8\textwidth]{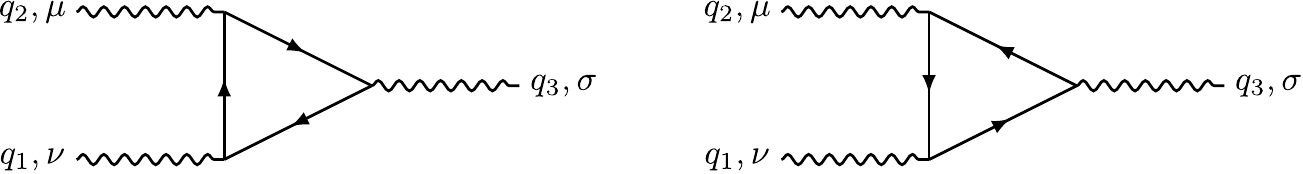}
\caption{\label{fig:Vert3F} Feynman diagrams for the $\gamma\gamma\gamma$ vertex at one-loop.}
\end{figure}
As expected, the contribution to the  $\gamma\gamma\gamma$ vertex from the diagrams in Figure \ref{fig:Vert3F} vanishes identically from the charge conjugation invariance of the theory.

\subsection{TT$\gamma$ vertex}

\begin{figure}[h]
\centering 
\includegraphics[width=.8\textwidth]{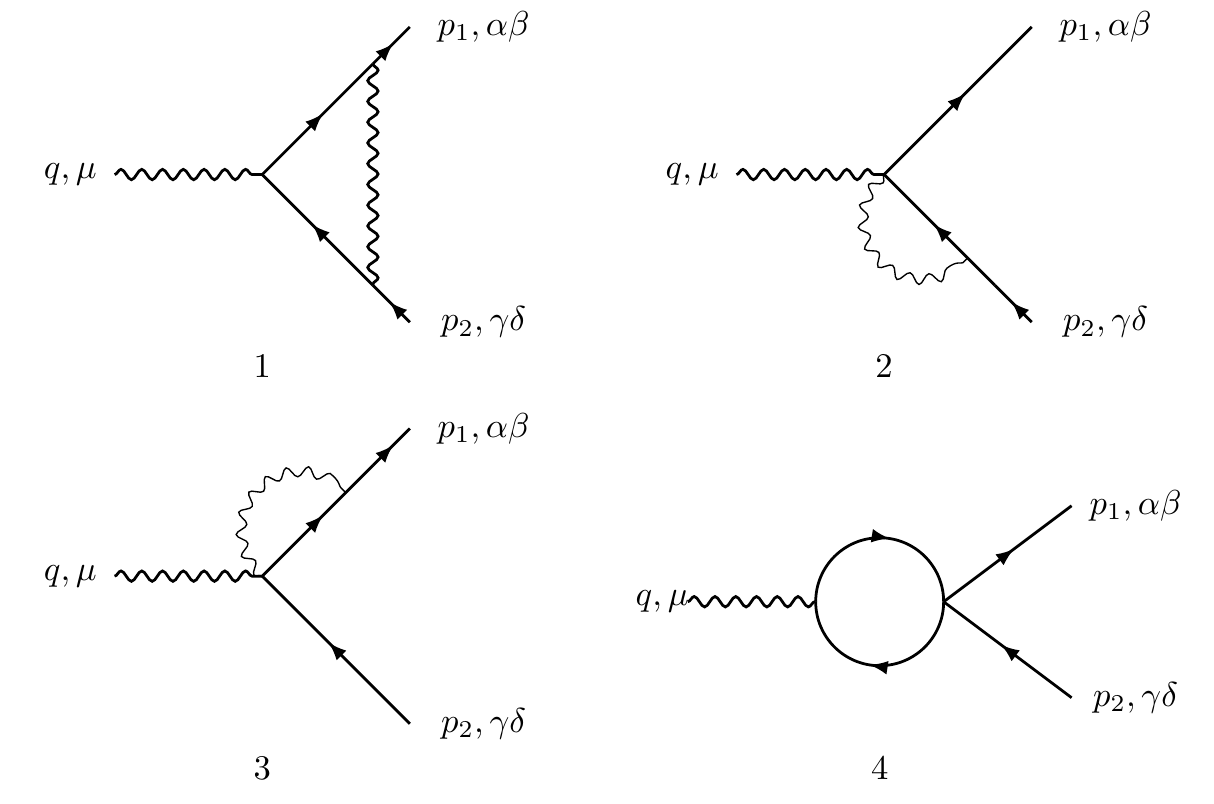}
\caption{\label{fig:Vert3} Feynman diagrams for the TT$\gamma$ vertex at one-loop.}
\end{figure}
The one-loop contribution to the TT$\gamma$ vertex comes from the four diagrams in Figure \ref{fig:Vert3}. Its divergent piece can be written as
\begin{eqnarray}
-ie\Gamma^{*\mu}_{\alpha\beta\gamma\delta}(-p_1-p_2,p_1,p_2)&=&-i\left[\frac{e^3 (\xi -3)}{16 \pi ^2 \epsilon }\right][T_{\mu\rho}p_{2}^{\rho}-T_{\rho\mu}p_{1}^{\rho}]_{\alpha\beta\gamma\delta}\\\nonumber
&&-eg\left[\frac{e^2 \left(g^2+2\right)+\lambda _1+\lambda _2+12 \lambda _3}{16 \pi ^2
   \epsilon }\right](p_{1}^{\rho}+p_{2}^{\rho})(M_{\mu\rho})_{\alpha\beta\gamma\delta},
\end{eqnarray}
and is canceled by the corresponding counterterm with
 \begin{eqnarray}
 &&\delta_{e}=-\frac{e^2 (\xi -3)}{16 \pi ^2 \epsilon },\label{deltae}\\
 &&\delta_{eg}=-\frac{e^2 \left(g^2+2\right)+\lambda _1+\lambda _2+12 \lambda _3}{16 \pi ^2
   \epsilon }.\label{deltaeg}
 \end{eqnarray}
 Notice that this result is consistent with the Ward identity
 \begin{equation}
\Gamma^{*\mu}(0,p,-p)=-\frac{\partial\Sigma^*(p)}{\partial p_\mu}.
\end{equation}
as $\delta_{e}=\delta_{2}$. Gauge invariance also fixes the counterterm involved in the finiteness of the TT$\gamma\gamma$ vertex, as Eq.(\ref{Ward2}) dictates that 
 \begin{eqnarray}
 &&\delta_{e2}=\delta_{e}=-\frac{e^2 (\xi -3)}{16 \pi ^2 \epsilon }.\label{deltae2}
 \end{eqnarray}

\subsection{TT$\gamma\gamma$ vertex}

\begin{figure}[h]
\centering 
\includegraphics[width=1\textwidth]{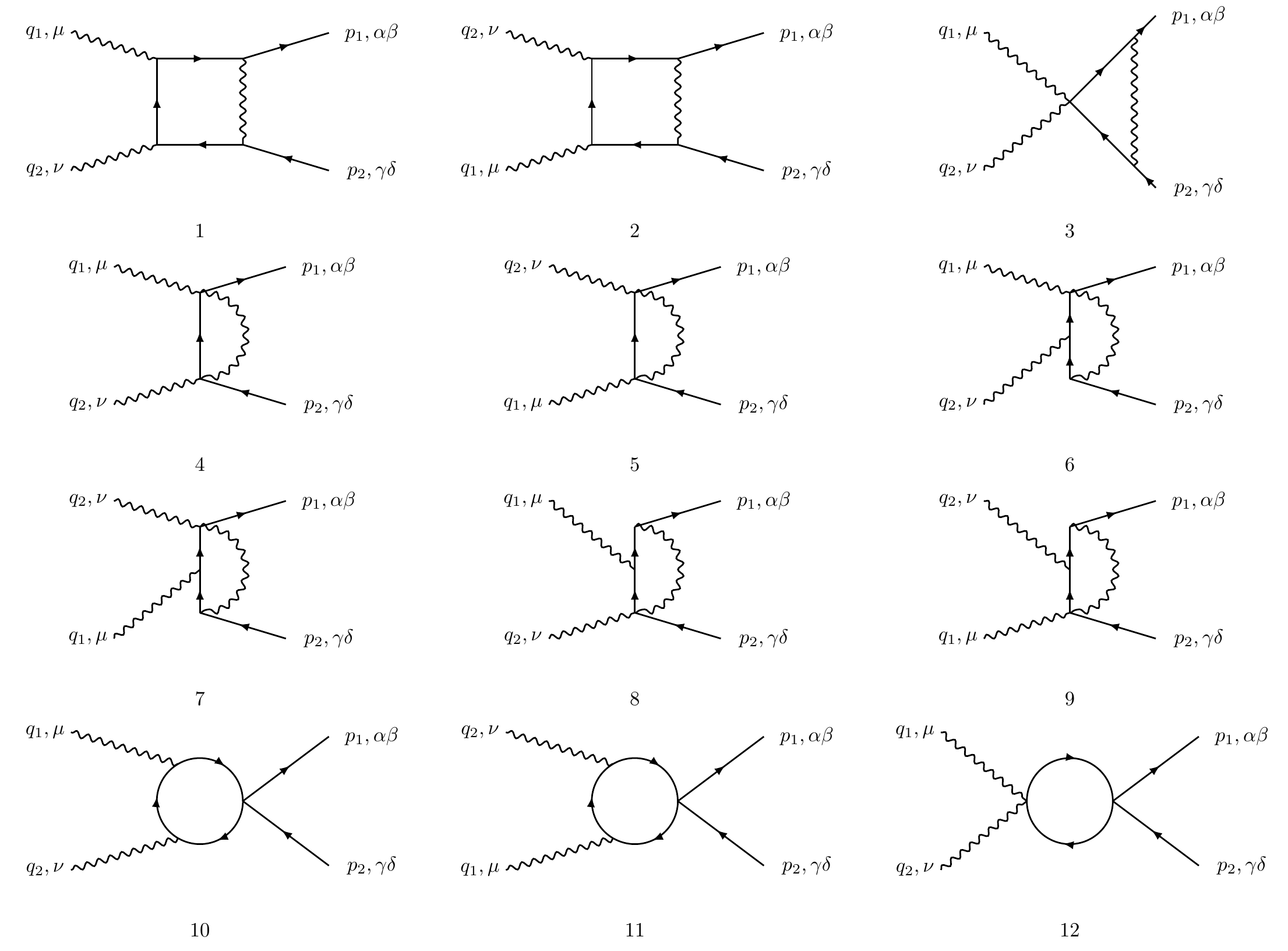}
\caption{\label{fig:TTgg} Feynman diagrams for the TT$\gamma\gamma$ vertex at one-loop.}
\end{figure}
There are 12 diagrams contributing to the TT$\gamma\gamma$ vertex at one-loop, as shown in Figure \ref{fig:TTgg}. The corresponding divergent piece is
\begin{equation}
ie^2\Gamma^{*\mu\nu}_{\alpha\beta\gamma\delta}=ie^2 \left[\frac{e^2 (\xi -3)}{8\pi^2 \epsilon }\right]1_{\alpha\beta\gamma\delta}g^{\mu\nu},
\end{equation}
and, as anticipated from the Ward identities, the full TT$\gamma\gamma$ vertex becomes finite with $\delta_{e2}$ given by Eq.(\ref{deltae2}).

\subsection{$\gamma\gamma\gamma\gamma$ vertex}

\begin{figure}[h]
\centering 
\includegraphics[width=1\textwidth]{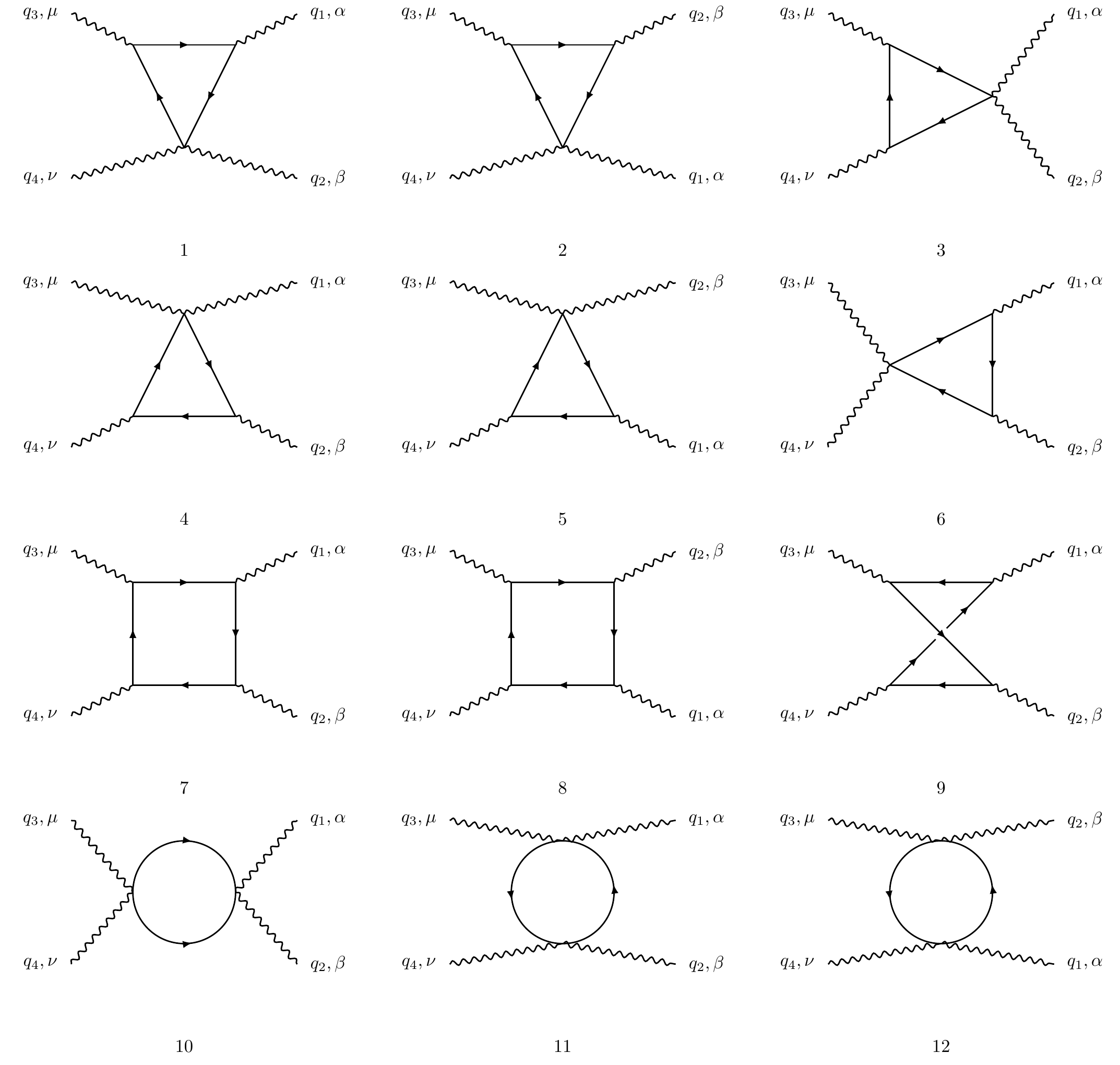}
\caption{\label{fig:Vert4F} Feynman diagrams for the $\gamma\gamma\gamma\gamma$ vertex at one-loop. There are $9$ additional diagrams obtained from diagrams $1-9$ reversing the arrow direction in the loop}
\end{figure}
The one-loop correction to the $\gamma\gamma\gamma\gamma$ vertex involves 21 diagrams, shown in Figure \ref{fig:Vert4F}, and there is no counterterm available to cancel a potential divergence in this case. By an explicit calculation, we have found that the divergent piece of the total amplitude vanishes exactly.

\subsection{TTTT vertex}

\begin{figure}[tp]
\centering 
\includegraphics[width=1\textwidth]{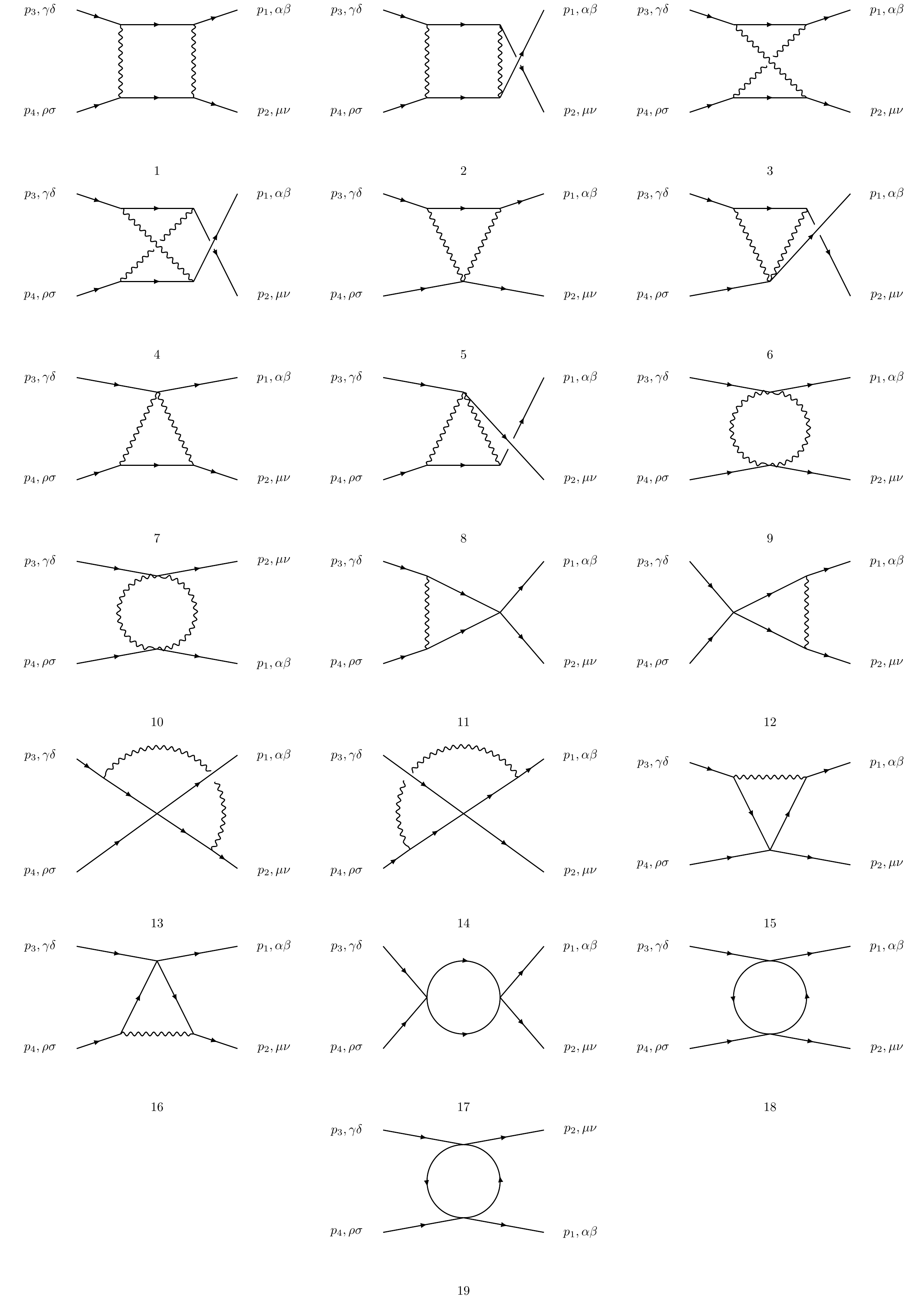}
\caption{\label{fig:TTTT} Feynman diagrams for the TTTT vertex at one-loop.}
\end{figure}

The last potentially divergent function is the TTTT vertex and there are 19 diagrams contributing to the total amplitude, as shown in Fig. \ref{fig:TTTT}. The divergent part of the TTTT vertex is 
\begin{eqnarray}
&&i\Lambda_{\alpha\beta\gamma\delta\mu\nu\rho\sigma}=\frac{1}{16\pi^2 \epsilon}\Big\{e^4 (3 g^4-8 g^2+6)+2 \lambda _1 \left(e^2 (2 g^2+\xi)+\lambda _2+8 \lambda _3+12 \lambda _4\right)\\ 
&&+11 \lambda _1^2+3 \lambda_2^2-8 \lambda_2 \lambda _4 \Big\}(1_{\alpha\beta\gamma\delta}1_{\mu\nu\rho\sigma}+1_{\alpha\beta\rho\sigma}1_{\mu\nu\gamma\delta})\nonumber\\
&&+\frac{1}{8\pi^2 \epsilon}\Big\{\lambda _2 \left(e^2 \left(2 g^2+\xi \right)+4 \lambda_1+8\lambda_3-8\lambda_4\right)+8 \left(\lambda _3-\lambda _4\right) \left(e^2 g^2+3\lambda _3-3 \lambda _4\right)\nonumber\\
&&+4 \lambda _2^2\Big\}(\chi_{\alpha\beta\gamma\delta}\chi_{\mu\nu\rho\sigma}+\chi_{\alpha\beta\rho\sigma}\chi_{\mu\nu\gamma\delta})\nonumber\\&&-\frac{1}{16\pi^2 \epsilon}\Big\{e^2 g^2 \left(\lambda _1+\lambda_2\right)+2 \lambda _3 \left(e^2 \xi +4 \lambda _1+4 \lambda _2\right)\nonumber\\
&&+8\lambda _3^2-24 \lambda _4^2\Big\}[M^{\kappa\lambda}_{\alpha\beta\gamma\delta}(M_{\kappa\lambda})_{\mu\nu\rho\sigma}+M^{\kappa\lambda}_{\alpha\beta\rho\sigma}(M_{\kappa\lambda})_{\mu\nu\gamma\delta}]\nonumber\\
&&-\frac{1}{64\pi^2 \epsilon}\Big\{e^4 g^4+8 \lambda _4 \left(e^2 \left(2 g^2-\xi \right)-4 \lambda _1+16
   \lambda _3-8 \lambda _4\right)\Big\}[S^{\kappa\lambda}_{\alpha\beta\gamma\delta}(S_{\kappa\lambda})_{\mu\nu\rho\sigma}+S^{\kappa\lambda}_{\alpha\beta\rho\sigma}(S_{\kappa\lambda})_{\mu\nu\gamma\delta}],\nonumber
\end{eqnarray}
and the corresponding counterterms that render the total amplitude finite are given in the MS scheme by
\begin{eqnarray}
&&\delta_{\lambda 1}=-\frac{1}{16 \pi ^2 \lambda _1 \epsilon }\Big\{e^4 (3 g^4-8 g^2+6)+2 \lambda _1 \left(e^2 (2 g^2+\xi)+\lambda _2+8 \lambda _3+12 \lambda _4\right)\nonumber\\ 
&&\qquad\qquad +16 \left(\lambda _4 (e^2 g^2+6 \lambda _3)+\lambda _3 (e^2 g^2+3 \lambda_3)+6 \lambda _4^2\right)\\ 
&&\qquad\qquad+11 \lambda _1^2+3 \lambda_2^2-8 \lambda_2 \lambda _4 \Big\},\nonumber \label{deltalambda1}\\
&&\delta_{\lambda 2}=-\frac{1}{8 \pi ^2 \lambda _2 \epsilon}\Big\{\lambda _2 \left(e^2 \left(2 g^2+\xi \right)+4 \lambda _1+8 \lambda _3-8\lambda_4\right)\\
&&\qquad\qquad +8 \left(\lambda _3-\lambda _4\right) \left(e^2 g^2+3\lambda _3-3 \lambda _4\right)+4 \lambda _2^2\Big\},\nonumber \label{deltalambda2}\\
&&\delta_{\lambda 3}=-\frac{1}{16 \pi ^2\lambda _3 \epsilon}\Big\{e^2 g^2 \left(\lambda _1+\lambda_2\right)+2 \lambda _3 \left(e^2 \xi +4  \lambda _1+4 \lambda _2\right)\\ 
&&\qquad\qquad +8 \lambda _3^2-24 \lambda _4^2\Big\},\nonumber \label{deltalambda3}\\
&&\delta_{\lambda 4}=\frac{1}{64 \pi ^2 \lambda _4 \epsilon }\Big\{e^4 g^4+8 \lambda _4 \left(e^2 \left(2 g^2-\xi \right)-4 \lambda _1+16
   \lambda _3-8 \lambda _4\right)\Big\}.\label{deltalambda4}
\end{eqnarray}

\subsection{Beta Functions}\label{betafsQED}
From the results obtained in eqs.(\ref{delta1},\ref{deltam},\ref{delta2},\ref{deltae},\ref{deltaeg},\ref{deltae2},\ref{deltalambda1},\ref{deltalambda2},\ref{deltalambda3},\ref{deltalambda4})
and the definitions in eqs.(\ref{deltas},\ref{Zs}), the relation between the bare and renormalized parameters of the model is
\begin{equation}
\begin{array}{c}
\begin{array}{ccccc}
 e_0=Z_1^{-\frac{1}{2}}Z_2^{-1}Z_e\mu^{\epsilon} e ,&\quad & e_0^2=Z_1^{-1} Z_2^{-1} Z_{e2} \mu^{2\epsilon}e^2, &\quad&
\lambda_{0j}=Z_{2}^{-2}Z_{\lambda_j}\mu^{2\epsilon}\lambda_{j},
\end{array}
\\
\begin{array}{ccc}
g_0=Z_e^{-1}Z_{eg}g,&\quad& m_0^2=Z_2^{-1}Z_mm^2,
\end{array}
\end{array}
\label{param}
\end{equation}
The renormalization constants in the MS subtraction scheme are
\begin{eqnarray}
&&Z_1=1+\frac{e^2(2g^{2}-1)}{8\pi^{2}\epsilon},\label{Z1}\\
&&Z_2=Z_{e2}=
Z_e=1-\frac{e^2 (\xi -3)}{16 \pi ^2 \epsilon },\\
&&Z_{\lambda_1}= 1-\frac{1}{16 \pi ^2 \lambda _1 \epsilon }\Big\{e^4 (3 g^4-8 g^2+6)+2 \lambda _1 \left(e^2 (2 g^2+\xi)+\lambda _2+8 \lambda _3+12 \lambda _4\right)\nonumber\\ 
&&\qquad\qquad +16 \left(\lambda _4 (e^2 g^2+6 \lambda _3)+\lambda _3 (e^2 g^2+3 \lambda_3)+6 \lambda _4^2\right)\\ 
&&\qquad\qquad+11 \lambda _1^2+3 \lambda_2^2-8 \lambda_2 \lambda _4 \Big\},\nonumber\\
&&Z_{\lambda_2}=1-\frac{1}{8 \pi ^2 \lambda _2 \epsilon}\Big\{\lambda _2 \left(e^2 \left(2 g^2+\xi \right)+4 \lambda _1+8 \lambda _3-8\lambda_4\right)\\
&&\qquad\qquad +8 \left(\lambda _3-\lambda _4\right) \left(e^2 g^2+3\lambda _3-3 \lambda _4\right)+4 \lambda _2^2\Big\},\nonumber\\
&&Z_{\lambda_3}=1 -\frac{1}{16 \pi ^2\lambda _3 \epsilon}\Big\{e^2 g^2 \left(\lambda _1+\lambda_2\right)+2 \lambda _3 \left(e^2 \xi +4  \lambda _1+4 \lambda _2\right)\\ 
&&\qquad\qquad +8 \lambda _3^2-24 \lambda _4^2\Big\},\nonumber\\
&&Z_{\lambda_4}=1+ \frac{1}{64 \pi ^2 \lambda _4 \epsilon }\Big\{e^4 g^4+8 \lambda _4 \left(e^2 \left(2 g^2-\xi \right)-4 \lambda _1+16
   \lambda _3-8 \lambda _4\right)\Big\},\\
&&Z_{eg}=Z_{e}+\delta_{g} =
1-\frac{1}{16 \pi^2 \epsilon }\Big\{e^2 (g^2+\xi -1)+\lambda _1+\lambda _2+12 \lambda _3\Big\},\\
&& Z_{m}=Z_{2}+\delta_m = 1-\frac{1}{16 \pi ^2 \epsilon}\Big\{e^2 \left(2 g^2+\xi\right)+7 \lambda _1+\lambda _2+8 \lambda_3+12\lambda_4\Big\}.\label{Zmf}
\end{eqnarray}
With the above results, the two different relations between $e_0$ and $e$ in eq.(\ref{param}) become
\begin{equation}
e_0= Z_1^{-1/2}\mu^{\epsilon} e. \label{renorcarga}
\end{equation}

From eqs. (\ref{param}-\ref{Zmf}) one can derive the relevant beta functions $\beta_\eta\equiv\mu\frac{\partial\eta}{\partial\mu}$ and anomalous dimensions $\gamma_m\equiv\frac{\mu}{m}\frac{\partial m}{\partial\mu}$ of the theory in the $\epsilon\to 0$ limit:
\begin{eqnarray}
&&\beta_e=\frac{e^3 \left(1-2 g^2\right)}{8 \pi ^2},\label{betafue}\\
&&\beta_{g}=-\frac{g \left[e^2 \left(g^2+2\right)+\lambda _1+\lambda _2+12 \lambda
   _3\right]}{8 \pi ^2},\label{funcbeg}\\
&&\beta_{\lambda_1}=\frac{1}{8 \pi ^2}\Bigg\{e^4 \left(-3 g^4+8 g^2-6\right)-2 \lambda _1 \left(e^2 \left(2
   g^2+3\right)+\lambda _2+8 \lambda _3+12 \lambda _4\right)\\&&\qquad\qquad -16 \left(\lambda _4
   \left(e^2 g^2+6 \lambda _3\right)+\lambda _3 \left(e^2 g^2+3 \lambda
   _3\right)+6 \lambda _4^2\right)-11 \lambda _1^2-3 \lambda _2^2+8 \lambda _2
   \lambda _4\Bigg\},\nonumber\\
&&\beta_{\lambda_2}=-\frac{1}{4 \pi ^2}\Big\{\lambda _2 \left(e^2 \left(2 g^2+3\right)+4 \lambda _1+8 \lambda _3-8\lambda _4\right)\\&&\qquad\qquad+8 \left(\lambda _3-\lambda _4\right) \left(e^2 g^2+3
   \lambda _3-3 \lambda _4\right)+4 \lambda _2^2\Big\},\nonumber\\
&&\beta_{\lambda_3}=-\frac{1}{8 \pi ^2}\Big\{e^2 g^2 \left(\lambda _1+\lambda _2\right)+2 \lambda _3 \left(3 e^2+4\lambda _1+4 \lambda _2\right)+8 \lambda _3^2-24 \lambda _4^2\Big\},\\
&&\beta_{\lambda_4}=\frac{1}{32 \pi ^2}\Big\{e^4 g^4+8 \lambda _4 \left(e^2 \left(2 g^2-3\right)-4 \lambda _1+16\lambda _3-8 \lambda _4\right)\Big\},\\
&&\gamma_m=-\frac{1}{16 \pi ^2}\Big\{e^2 \left(2 g^2+3\right)+7 \lambda _1+\lambda _2+8 \lambda _3+12 \lambda_4\Big\}.\label{gammam}
\end{eqnarray}

We conclude this section with a short discussion of  some of the possible scenarios of the theory. There is a trivial fixed point for the beta functions of the theory when $g=0$, $\lambda_2=0$, $\lambda_3=0$ and $\lambda_4=0$. This fixed point corresponds to the limit in which each component of the tensor $B^{\mu\nu}$ behaves as a complex scalar field in a $\lambda \phi^4$ theory with $\lambda_1=-\lambda/2$.
On the other hand, the $\beta_{\lambda_i}$
are all nonzero for any non-vanishing real value of the gyromagnetic factor $g$, even if all self interactions are set to $\lambda_i=0$, $i=1,\dots,4$. This means that, oppositely to the spin $1/2$ case studied in  \cite{VaqueraAraujo:2012qa},  pure electrodynamics for matter fields of spin $1$ is not viable for $g\neq0$, as  self interactions are necessary to make the theory renormalizable.
Finally, turning off the electromagnetic interactions by taking $e=0$ and $g=0$, the theory reduces to a renormalizable model of pure self-interacting terms  for the tensor fields, with
\begin{eqnarray}
&&\beta_{\lambda_1}=-\frac{1}{8 \pi ^2}\Big\{11 \lambda _1^2+2\lambda_1 \left(\lambda _2+8 \lambda_3+12 \lambda _4\right)+3 \lambda _2^2+48\lambda _3^2\\
   &&\qquad\qquad -8 \lambda _2 \lambda _4+96 \lambda _4\left(\lambda _3+\lambda _4\right)\Big\},\nonumber\\
&&\beta_{\lambda_2}=-\frac{1}{\pi ^2}\Big\{\lambda _2^2+\lambda _1 \lambda _2+2\lambda_2 \left(\lambda _3-\lambda _4\right)+6\left(\lambda _3-\lambda _4\right){}^2\Big\},\\
&&\beta_{\lambda_3}=\frac{1}{\pi ^2}\Big\{3 \lambda _4^2-\lambda _3 \left(\lambda
   _1+\lambda _2+\lambda _3\right)\Big\},\\
&&\beta_{\lambda_4}=-\frac{1}{\pi ^2}\Big\{\lambda _4 \left(\lambda _1-4 \lambda _3+2
   \lambda _4\right)\Big\},\\
   &&\gamma_m=-\frac{1}{16 \pi ^2}\Big\{7 \lambda _1+\lambda _2+8 \lambda _3+12 \lambda_4\Big\}.\label{gammam.ms}
\end{eqnarray}

\section{Summary and conclusions}\label{conc}
In this work, we have studied the one-loop renormalization of the electrodynamics of fields transforming under the $(1,0)\oplus(1,0)$ representation of the HLG in the Poincar\'e projector formalism. The analysis has been done in an arbitrary covariant gauge, with arbitrary gyromagnetic factor and including all the independent parity conserving self-interactions.  The main conclusion of the work is that the theory is renormalizable for any value of the gyromagnetic factor, displaying a rich set of renormalization group equations. In contrast to the analogous spin 1/2 case studied in \cite{VaqueraAraujo:2012qa}, there is no non-trivial finite value for the gyromagnetic factor that allows the existence of a pure electrodynamics without the inclusion of self interactions.

\acknowledgments

This work is dedicated to the memory of Sitka de los Ríos. A.R-A. acknowledges support by CONACyT. C.A.V-A. is supported by the Mexican C\'atedras CONACyT project 749 and SNI 58928.


\end{document}